\documentclass[conference]{IEEEtran}
\usepackage{graphicx}
\usepackage{algorithm,algorithmic}

\ifCLASSINFOpdf
\else
\fi
\begin{document}
%
\title{System Identification Using Reweighted Zero Attracting Least Absolute Deviation Algorithms}

\author{\IEEEauthorblockN{Fuxi Wen}
\IEEEauthorblockA{School of Electrical and Electronic Engineering
 \\Nanyang Technological University\\Singapore
\\
Email: wenfuxi@ntu.edu.sg}
}


%


\maketitle

\begin{abstract}
In this paper,  the $l_1$ norm penalty on the filter coefficients is incorporated in the least mean absolute deviation (LAD) algorithm to improve the performance of the LAD algorithm.  The performance of LAD, zero-attracting LAD (ZA-LAD)  and reweighted  zero-attracting LAD (RZA-LAD) are evaluated for linear time varying system identification under the non-Gaussian ($\alpha$-stable)  noise environments. Effectiveness of the ZA-LAD type algorithms is demonstrated through computer simulations.
\end{abstract}


\section{Introduction}
\label{S1}
Adaptive  filters have been widely used in active noise control, channel equalization, adaptive inverse control, echo cancellation, noise cancellation, linear prediction and system identification.  Least mean squares (LMS), recursive least squares (RLS) and their variations are the widely used  adaptive algorithms \cite{ref1b}-\cite{ref4b}. The least mean squares (LMS) algorithm adjusts the filter coefficients to minimize the least mean squares of the error signal. Compared to recursive least squares (RLS) algorithm, the LMS algorithm has a slower convergence speed, however it does not involve any matrix operations. Therefore, the LMS algorithm requires fewer computational resources and memory than the RLS algorithm. The implementation of the LMS algorithm also is less complicated than the RLS algorithm. 

In many applications, the unknown system response can be assumed to be sparse, which means only a small fraction of the coefficients are different from zero. For example,  sparse echo cancellation \cite{ref20}, internet  telephony, sparse nonlinear channel estimation \cite{ref33}, sparse system estimation \cite{ref34}-\cite{ref35}. Zero-attracting filters exploiting the sparse nature of the system can improve the  performance of the adaptive filter. A similar idea is used in linear prediction of speech in \cite{ref36} to obtain better coding properties by enhancing the sparsity of residual and predictor.  

Motivated by Lasso \cite{ref10} and recent progress in compressive
sensing \cite{ref11},\cite{ref12}, the zero-attracting LMS (ZA-LMS) algorithm and reweighted  zero-attracting LMS (RZA-LMS) were proposed in \cite{ref24},  for sparse system identification.  ZA-LMS is a combination of LMS algorithm with $l_1$-norm penalty of the coefficient vector. The reweighted $l_1$-norm minimization algorithm is first proposed in \cite{refrl} to enhance the sparsity of the system. Convergence analysis of ZA-LMS is given in \cite{ref25}. To reduce the eigenvalue spread of the input signal correlation matrix, ZA-LMS and RZA-LMS are extended to the transform domain in \cite{ref26}. 

The LMS type algorithms generally provide a more accurate solution, with less mis-adjustment when the noise is Gaussian, however, they are very sensitive to the outliers. On the other hand, though the convergence rate of the LAD type algorithms is low,  they are robust to the outliers, such as the $\alpha$-stable noise. The $\alpha$-stable noise model is a more generalized model \cite{ref6lmp}-\cite{ref7lmpsi}, such that the Gaussian model can be seen as an special case of $\alpha$-stable model by setting the characteristic exponent $\alpha = 2$.

In this paper, the performance of the least mean absolute deviation (LAD) \cite{ref6lmp}, zero-attracting LAD (ZA-LAD) \cite{ref32} and reweighted  zero-attracting LAD (RZA-LAD) algorithms are evaluated for linear time varying system identification under the non-Gaussian ($\alpha$-stable)  noise environments. 

The rest of the paper is organized as follows. The LAD, ZA-LAD and RZA-LAD algorithms are given in Section II. A brief discussion about the parameter selection is provided in Section III. In Section IV, we evaluate the performance of the three algorithms for linear time varying system identification under Gaussian and non-Gaussian noise environments.  Conclusions of our work and  some further research directions are provided in Section V.

\section {The LAD, ZA-LAD and RZA-LAD algorithms}
\label{S2}
Notations:  $\mathbf{x}(n)$ is the system input signal at time $n$, $\mathbf{w}(n)$ is the coefficient vector and  $d(n)$ is the desire signal at time $n$, The superscripts $(\cdot)^T$ denotes the transpose and $\textrm{sgn}(\cdot)$ denotes the sign function. $|\cdot|$ and $||\cdot||_1$ denote the absolute value and  $l_1$-norm, respectively.
\subsection {The LAD algorithm}
\label{S2-1}
The output of the LAD algorithm is given by
\begin{equation}
\label{eq_1}
y_1(n) = \mathbf{w}_1^T(n)\mathbf{x}(n).
\end{equation}
The cost function is formulated as
\setlength\arraycolsep{0.1em}
\begin{eqnarray}
\label{eq_2}
J_1\big(\mathbf{w}_1(n)\big)  &=&   \left|d(n) - y_1(n)  \right| \nonumber\\
&=& \left| e_1(n) \right|,
\end{eqnarray}
where $e_1(n) = d(n) - y_1(n)$ is the error signal.

Using the stochastic gradient approach, the filter weights are estimated iteratively by
\begin{equation}
\label{eq_3}
\mathbf{w}_1(n+1)=\mathbf{w}_1(n) - \mu_1\frac{ \partial J_1\big(\mathbf{w}_1(n)\big)}{ \partial \mathbf{w}_1(n)},
\end{equation}
where 
\begin{equation}
\label{eq_4}
\frac{ \partial J_1\big(\mathbf{w}_1(n)\big)}{ \partial \mathbf{w}_1(n)} = -\textrm{sgn} \big(e_1(n)\big)\mathbf{x}(n).
\end{equation}
Substituting (\ref{eq_4}) into (\ref{eq_3}), we obtain the updating equation for the LAD algorithm:
\begin{equation}
\label{eq_5}
\mathbf{w}_1(n+1)=\mathbf{w}_1(n)+ \mu_1 \textrm{sgn} \big(e_1(n)\big)\mathbf{x}(n),
\end{equation}
where $\mu_1$ is the step size that should be chosen carefully to ensure convergence.

The LAD algorithm is summarized in Algorithm \ref{alg1}.
\begin{algorithm}[!h]
\caption{LAD}
\label{alg1}
\begin{algorithmic}
\STATE   Inputs:  $\mu_1$
\STATE  Initialization:  $\mathbf{w}_1(0)=0$
\FOR{$n = 0, 1, 2, \cdots $} 
\STATE $y_1(n)  =  \mathbf{w}^T_1(n)\mathbf{x}(n)$
\STATE $e_1(n)  = d(n) - y_1(n)$
\STATE $\mathbf{w}_1(n+1) = \mathbf{w}_1(n)+\mu_1\textrm{sgn}\big(e_1(n)\big)\mathbf{x}(n)$
\ENDFOR
\end{algorithmic}
\end{algorithm}
\subsection {The ZA-LAD algorithm}
\label{S2-2}
The output of the ZA-LAD algorithm is 
\begin{equation}
\label{eq_6}
y_2(n) = \mathbf{w}_2^T(n)\mathbf{x}(n).
\end{equation}
For the ZA-LAD algorithm, $l_1$ norm penalty is used to explore the sparse nature of the filter coefficients. The cost function is formulated as
\setlength\arraycolsep{0.1em}
\begin{eqnarray}
\label{eq_7}
J_2\big(\mathbf{w}_2(n)\big)  &=&   \left|d(n) - y_2(n)  \right|  + \alpha_2 \left|\left|\mathbf{w}_2(n)\right|\right|_1 \nonumber\\
&=&   \left| e_2(n) \right|  + \alpha_2\left|\left|\mathbf{w}_2(n)\right|\right|_1,
\end{eqnarray}
where $e_2(n) = d(n) - y_2(n)$ is the error signal, $\alpha_2$ is the regularization parameter.

Using the stochastic gradient approach, the filter weights are estimated iteratively by
\begin{equation}
\label{eq_8}
\mathbf{w}_2(n+1)=\mathbf{w}_2(n) - \mu_2\frac{ \partial J_2\big(\mathbf{w}_2(n)\big)}{ \partial \mathbf{w}_2(n)},
\end{equation}
where 
\begin{equation}
\label{eq_9}
\frac{ \partial J_2\big(\mathbf{w}_2(n)\big)}{ \partial \mathbf{w}_2(n)} = -\textrm{sgn} \big(e_2(n)\big)\mathbf{x}(n)  + \alpha_2  \textrm{sgn}  \big(\mathbf{w}_2(n)\big).
\end{equation}
Substituting (\ref{eq_9}) into (\ref{eq_8}), we obtain the updating equation for the ZA-LAD algorithm:
\begin{equation}
\label{eq_10}
\mathbf{w}_2(n+1)=\mathbf{w}_2(n)+ \mu_2 \textrm{sgn} \big(e_2(n)\big)\mathbf{x}(n)  - r_2  \textrm{sgn}  \big(\mathbf{w}_2(n)\big),
\end{equation}
where $\mu_2$ is the stepsize and $r_2=\mu_2\alpha_2$.

The ZA-LAD algorithm is summarized in Algorithm \ref{alg2}.
\begin{algorithm}[!h]
\caption{ZA-LAD}
\label{alg2}
\begin{algorithmic}
\STATE   Inputs:  $ \mu_2,  r_2$
\STATE  Initialization:  $\mathbf{w}_2(0)=0$
\FOR{$n = 0, 1, 2, \cdots $} 
\STATE $y_2(n)  =  \mathbf{w}^T_2(n)\mathbf{x}(n)$
\STATE $e_2(n)  = d(n) - y_2(n)$
\STATE $\mathbf{w}_2(n+1) = \mathbf{w}_2(n)+\mu_2\textrm{sgn}\big(e_2(n)\big)\mathbf{x}(n)-r_2\textrm{sgn}\big(\mathbf{w}_2(n)\big) $
\ENDFOR
\end{algorithmic}
\end{algorithm}
\subsection {The RZA-LAD algorithm}
\label{S2-3}
The output of the RZA-LAD algorithm is 
\begin{equation}
\label{eq_11}
y_3(n) = \mathbf{w}_3^T(n)\mathbf{x}(n).
\end{equation}

Compared with $l_1$ norm, log-sum penalty function is a better approximation for the sparsity of the filter coefficients. For the RZA-LAD algorithm, log-sum penalty function is used as the coefficient penalty function. The cost function is formulated as
\setlength\arraycolsep{0.1em}
\begin{eqnarray}
\label{eq_12}
J_3\big(\mathbf{w}_3(n)\big)  &=& \left|d(n) - y_3(n)  \right| + \alpha_3 \sum_{m=1}^{M} \log \big(1 + \varepsilon_3 |[\mathbf{w}_3(n)]_m|\big) \nonumber\\
&=& \left| e_3(n) \right| + \alpha_3 \sum_{m=1}^{M} \log \big(1 + \varepsilon_3 |[\mathbf{w}_3(n)]_m|\big),
\end{eqnarray}
where $e_3(n) = d(n) - y_3(n)$ is the error signal, $\alpha_3$ is the regularization parameter, $\varepsilon_3$ is a positive number and $M$ is the number of elements in weight vector $\mathbf{w}_3(n)$.

Using the stochastic gradient approach, the filter weights are estimated iteratively by
\begin{equation}
\label{eq_13}
\mathbf{w}_3(n+1)=\mathbf{w}_3(n) - \mu_3\frac{ \partial J_3\big(\mathbf{w}_3(n)\big)}{ \partial \mathbf{w}_3(n)},
\end{equation}
and 
\begin{equation}
\label{eq_14}
\frac{ \partial J_3\big(\mathbf{w}_3(n)\big)}{ \partial \mathbf{w}_3(n)} = -\textrm{sgn} \big(e_3(n)\big)\mathbf{x}(n)  +  \alpha_3\varepsilon_3\mathbf{g}(n).
\end{equation}
where $\mathbf{g}(n) \in \mathcal{R}^{M \times 1}$ and the $m$th element of $\mathbf{g}(n)$ is given by
\begin{equation}
\label{eq_15A}
\left[ \mathbf{g}(n) \right]_m =  \frac{ \textrm{sgn} \big([\mathbf{w}_3(n)]_m\big)}{1+\varepsilon_3 \left|[\mathbf{w}_3(n)]_m \right|},\; 1 \leq m \leq M.
\end{equation}
Substituting (\ref{eq_14}) into (\ref{eq_13}), we obtain the updating equation for the RZA-LAD algorithm.
The updating equation is given by
\begin{equation}
\label{eq_15}
\mathbf{w}_3(n+1) = \mathbf{w}_3(n) + \mu_3\textrm{sgn} \big(e_3(n)\big)\mathbf{x}(n)  - r_3\mathbf{g}(n),
\end{equation}
where $\mu_3$ is the stepsize and $r_3=\mu_3\alpha_3\varepsilon_3$ and $\mathbf{g}(n)$ is given in (\ref{eq_15A}).

The RZA-LAD algorithm is summarized in Algorithm \ref{alg3}.
\newpage
\begin{algorithm}[!h]
\caption{RZA-LAD}
\label{alg3}
\begin{algorithmic}
\STATE   Inputs:  $\mu_3,  r_3, \varepsilon_3$
\STATE  Initialization:  $\mathbf{w}_3(0)=0$
\FOR{$n = 0, 1, 2, \cdots $} 
\STATE $y_3(n)  =  \mathbf{w}^T_3(n)\mathbf{x}(n)$
\STATE $e_3(n)  = d(n) - y_3(n)$
\STATE $\mathbf{w}_3(n+1) = \mathbf{w}_3(n) + \mu_3\textrm{sgn} \big(e_3(n)\big)\mathbf{x}(n)  - r_3\mathbf{g}(n)$
\ENDFOR
\end{algorithmic}
\end{algorithm}
\section{Parameter Selection - A Brief Discussion}
\label{S3}
The ZA-LAD and RZA-LAD algorithms are regularization based adaptive algorithms.
For the regularization based adaptive algorithms, the cost function is defined by combining the the $l_p$ norm of the error signal with the $l_q$ norm penalty of the coefficient vector.

\subsection{The choice of $r$ ($r_2$ or $r_3$ in Algorithm \ref{alg2} or Algorithm \ref{alg3})}
\label{S3-1}
Regularization plays a fundamental role in adaptive filtering, however the better performance is not obtained if the regularization parameter is not 
chosen properly. According to  (\ref{eq_10}) and (\ref{eq_15}), the parameter $r$ denotes the importance of the $l_1$ norm term or the intensity of attraction. So a large $r$ results in a faster convergence since the intensity of attraction increases as $r$ increases. However, steady-state misalignment increases as $r$ increases. Therefore, the parameter $r$ are determined by the trade-off between adaptation speed and adaptation quality in particular applications. 

One possible way to find the optimal regularization parameter $\alpha$ for four LMS-type algorithms is given in \cite{ref22}. The performance of the regularization based algorithms may be improved by using an adaptive regularization factor. To verify the performance of the LAD type algorithms, a fixed regularization factor is used.

\subsection{The choice of $\varepsilon$}
\label{S3-2}
We introduce the parameter $\varepsilon < 0$ in (\ref{eq_12}) in order to provide stability and
to ensure that a zero-valued component in $\mathbf{w}(n)$ does not strictly prohibit a
nonzero estimate at the next step. As empirically demonstrated in Section \ref{S4-1}, the RZA-LAD algorithm is robust to the choice of $\varepsilon$.
As mentioned in \cite{refrl}, $\varepsilon$ should be set slightly smaller than the expected nonzero magnitudes of $\mathbf{w}(n)$.
\section{Simulations}
\label{S4}
In this section, the performance of the proposed method is assessed via computer simulations. Two experiments are designed to demonstrate the steady-state performance, convergence rate, and tracking ability of theLAD, ZA-LAD, RZA-LAD, LMS, ZA-LMS and RZA-LAD algorithms. For comparison purposes, we also implement the LMS, RZA-LMS and ZA-LMS algorithms. Mean square deviation (MSD) is taken as a metric, which is defined as 
\begin{equation}
\label{eq_16}
MSD = \frac{1}{K} \sum_{k=1}^{K} || \mathbf{w}^{(k)} - \mathbf{w}_o||_2^2,
\end{equation}
where $\mathbf{w}^{(k)}$ is the coefficient estimated in the $k$th  independent trial and $\mathbf{w}_o$ is the real coefficient of the system.

The noise is non-Gaussian with an $\alpha$-stable distribution. The Matlab $\alpha$-stable distribution toolbox written by Mark S. Veillette is used to generate the noise with $\alpha$-stable distribution. The characteristic exponent $\alpha = 1.2$, the symmetry parameter $\beta = 0$, the location parameter $\gamma = 0$ and the dispersion (also called the scale) $\delta = 1$. 

The six filters (LAD, ZA-LAD, RZA-LAD, LMS, ZA-LMS and RZA-LAD) are run 500 times independently. The parameters are  given in Table \ref{table1}.
\begin{table}[!h]
\caption{Simulation parameters}
\label{table1}
\centering
\begin{tabular}{ccccc}
\hline
Algorithms & $\mu$ & $r$ & $\varepsilon$ \\
\hline
LAD & $\mu_1 = 5\times 10^{-3}$&  & \\
ZA-LAD & $\mu_2 = 5\times 10^{-3}$ &$r_2=1.5\times 10^{-4}$ & \\
RZA-LAD &$\mu_3 = 5\times 10^{-3}$& $r_3=1.5\times 10^{-3}$ &$\varepsilon_3=1\times 10^{-2}$\\
LMS & $\mu_4 = 5\times 10^{-3}$&  & \\
ZA-LMS & $\mu_5 = 5\times 10^{-3}$ &$r_5=1.5\times 10^{-4}$ & \\
RZA-LMS &$\mu_6 = 5\times 10^{-3}$& $r_6=1.5\times 10^{-3}$ &$\varepsilon_6=1\times 10^{-2}$\\
 \hline
\end{tabular}
\end{table}
\subsection {Example 1: Robustness of parameter $\varepsilon$}
\label{S4-1}
\begin{figure}[!ht]
\centering
\includegraphics[width=0.5\textwidth]{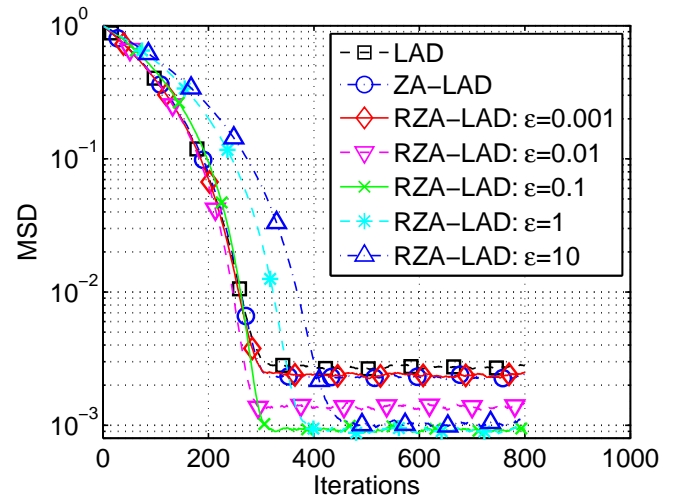}
\caption{Robustness of parameter $\varepsilon$. The input signal is Gaussian, noise is $\alpha$  stable, characteristic exponent  $\alpha  = 1.2$, GSNR = 10dB, 500 independent trials.}
\label{fig1_r}
\end{figure}

To verify the effect of parameter $\varepsilon$, the following experiment is considered.  A linear time invariant system with 16 coefficients is considered. Initially, we set the $5th$ tap with value 1 and the others to zero, making the system have a sparsity of 1/16. The performance of LAD, ZA-LAD and RZA-LAD with different  $\varepsilon$  is compared when the input signal is a white Gaussian random sequence with variance
of 1 and the generalized signal-to-noise ratio (GSNR) as defined in \cite{refgsnr} is 10 $dB$. The average MSD is shown in Fig.\ref{fig1_r}. 

From Fig.\ref{fig1_r}, we can see that, the performance of ZA-LAD and RZA-LAD is similar when $\varepsilon=0.001$ is used. It is because for a small $\varepsilon$, $\varepsilon \left|\mathbf{w}_3(n) \right| \approx 0$. The update functions for ZA-LAD (\ref{eq_10}) and RZA-LAD  (\ref{eq_15})  are equivalent, then the RZA-LAD algorithm reduces to the conventional ZA-LAD algorithm.
Compared with LAD and ZA-LAD, better steady-state performance is obtained for the RZA-LAD algorithm  when $\varepsilon=0.01, 0.1, 1$ or $10$. Another observation is that, the  convergence rate of RZA-LAD is slower when a larger $\varepsilon$ is used. It is because, $\varepsilon$ is in the denominator part of the third term of the RZA-LAD update function (\ref{eq_15}). A larger $\varepsilon$ will make the step-size in the third term of (\ref{eq_15}) smaller. The convergence rate is slower due to the relatively smaller step-size. 
\subsection {Example 2: Gaussian Input}
\label{S4-2}
In the second experiment, similar to the simulation setup used in \cite{ref25}, a linear time varying system with 16 coefficients is considered. Initially, we set the $5th$ tap with value 1 and the others to zero, making the system have a sparsity of 1/16. After 3000 iterations, all the odd taps are set to 1, while all the even taps remains
to be zero, i.e., a sparsity of 8/16. After 6000 iterations all the even taps are set with value -1 while all the odd taps are maintained to be
1, leaving a completely non-sparse system. 

\begin{figure}[!ht]
\centering
\includegraphics[width=0.5\textwidth]{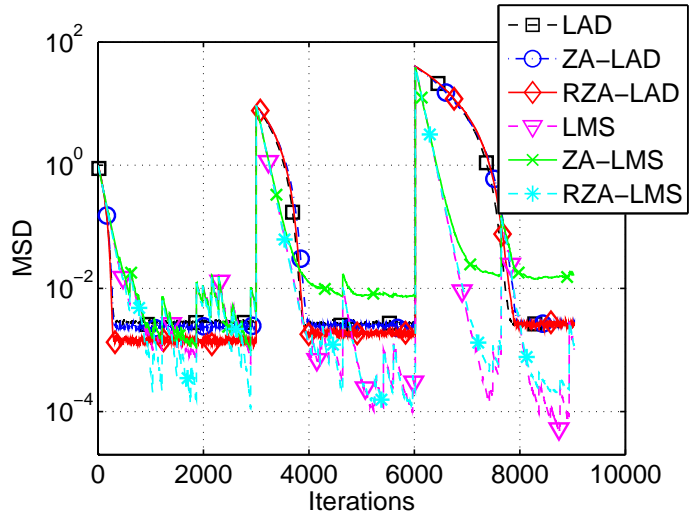}
\caption{Both of the input signal and noise are Gaussian, GSNR = 10dB, 500 independent trials.}
\label{fig1_g}
\end{figure}

The performance of LAD, ZA-LAD, RZA-LAD, LMS, ZA-LMS and RZA-LMS is compared when the input signal is a white Gaussian random sequence with variance
of 1 and the generalized signal-to-noise ratio (GSNR) is 10 $dB$. The average MSD is shown in Fig.\ref{fig1_g}. 

From Fig.\ref{fig1_g}, we can see that, when the system is very sparse (before the 3000$th$ iterations),  RZA-LAD yields fastest convergence rate and best steady-state performance compared with LAD and ZA-LAD. And the MSD of the LAD  and ZA-LAD algorithms are similar. After the 3000$th$ iteration, as the number of non-zero efficients increases to eight, a similar performance is obtained for LAD and ZA-LAD, while the RZA-LAD algorithm maintains the best performance among the three algorithms. After 6000$th$ iterations, the system is completely non-sparse, the performance of all the three algorithms are similar.
Since the impulsive noise is considered, all the three LMS type algorithms are not converged.

\subsection {Example 3: Non-Gaussian Input}
\label{S4-3}
The system in the second experiment is the same as the second one, except the switching times are set to the 20000$th$ iteration and 40000$th$ iteration. 

\begin{figure}[!ht]
\centering
\includegraphics[width=0.5\textwidth]{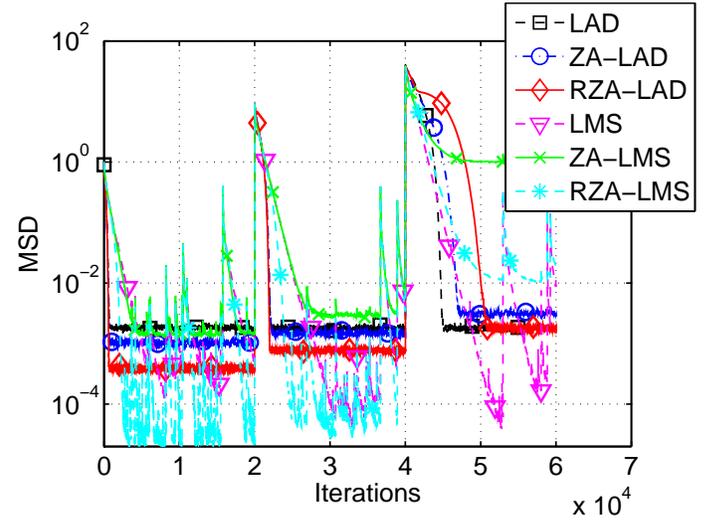}
\caption{The input signal is Gaussian, noise is $\alpha$  stable, characteristic exponent  $\alpha  = 1.2$, GSNR = 10dB, 500 independent trials.}
\label{fig2_ng}
\end{figure}

The performance of LAD, ZA-LAD, RZA-LAD, LMS, ZA-LMS and RZA-LMS is compared when the input signal $x(n)$ is now a correlated signal generated by $x(n)=0.8x(n-1)+u(n)$, where $u(n)$ is a white Gaussian noise. The average MSD is shown in Fig.\ref{fig2_ng}. 

From Fig.\ref{fig2_ng} we can see that, when the system is very sparse (before the 20000$th$ iterations), both  ZA-LAD and RZA-LAD yield faster convergence rate and better steady-state performance than LAD. And the MSD of the RZA-LAD algorithm is lower than the ZA-LAD algorithm. After the 20000$th$ iteration, as the number of non-zero efficients increases to eight, the performance of the ZA-LAD deteriorated, a similar performance is obtained for LAD and ZA-LAD, while the RZA-LAD algorithm maintains the best performance among the three algorithms. After th 40000$th$ iteration, the system is completely non-sparse, the performance of the LAD and RZA-LAD algorithms are similar.
The convergence rate of the RZA-LAD algorithm is low compared with LAD.
All the three LMS type algorithms are still not converged due to the impulsive noise.
\section{Conclusion}
In this paper, the performance of LAD,  ZA-LAD and  RZA-LAD are evaluated for linear time varying system identification under the Gaussian and non-Gaussian ($\alpha$-stable)  noise environments. The coefficients are updated using fixed stepsize. 

Better performance is obtained for the ZA-LAD type algorithms by exploiting the sparse nature of the system. RZA-LAD performs best among all the three algorithms, even when the system is non-sparse. Furthermore, the ZA-LAD type algorithms are robust to the non-Gaussian noise with $\alpha$-stable distribution. 

There are two potential further research directions, the first one is evaluating the performance of ZA-LAD type algorithms using variable stepsize and automatically adapted regularization parameters to ensure reliable performance across a broad array of signals.
The second one is extending the ZA-LAD type algorithms to nonlinear system such as nonlinear system identification using Volterra filter.




\end{document}